\documentclass[sigplan,10pt]{acmart}
\renewcommand\footnotetextcopyrightpermission[1]{}
\AtBeginDocument{%
  }

\setcopyright{acmlicensed}
\copyrightyear{2026}
\acmYear{2026}
\acmDOI{XXXXXXX.XXXXXXX}
\acmConference[Conf]{Conference}{MM YYYY}{City, Country}
\acmISBN{aaa}

\usepackage[linesnumbered, vlined, ruled]{algorithm2e}
\SetAlgorithmName{Procedure}{Procedure}{List of Procedures}
\SetArgSty{textrm} 

\SetCommentSty{mycommfont}
\usepackage{soul} 

\newcommand*\circled[1]{%
  \raisebox{-5pt}{%
    \setlength{\unitlength}{1pt}%
    \setlength{\fboxsep}{1pt}
    \setlength{\fboxrule}{1pt}
    \begin{picture}(14,14)
      \linethickness{1pt}
      \put(7,7){\circle{12}}
      \put(7,7){\makebox(0,0){#1}}
    \end{picture}%
  }%
}

\usepackage{listings}
\lstdefinestyle{Cstyle}{
  basicstyle=\ttfamily,
  language=C,
  keywordstyle=\bfseries,
  keywords={can_execute, can_read, can_write, can_call, cmpt_id, execution_context},
  commentstyle=\color{gray}\itshape\small,
  breaklines=true,
  breakatwhitespace=true,
  basewidth=0.8em,
  frame=tb, 
  columns=fullflexible
}

\graphicspath{{fig}{./fig/}}

\setstcolor{red} 

\newcommand{\sys}{Pomegranate}

\begin{document}

\title{\sys{}: A Lightweight Compartmentalization Architecture using Virtualization Extensions}

\author{Shriram Raja}
\authornote{Both authors contributed equally.}
\email{shriramr@bu.edu}
\orcid{0000-0002-3123-3486}
\affiliation{%
  \institution{Boston University}
  \city{Boston}
  \state{Massachusetts}
  \country{USA}
}

\author{Zhiyuan Ruan}
\authornotemark[1]
\email{zruan@bu.edu}
\orcid{0009-0008-2616-3162}
\affiliation{%
  \institution{Boston University}
  \city{Boston}
  \state{Massachusetts}
  \country{USA}
}

\author{Richard West}
\email{richwest@cs.bu.edu}
\orcid{0000-0001-5100-0666}
\affiliation{%
  \institution{Boston University}
  \city{Boston}
  \state{Massachusetts}
  \country{USA}
}


\begin{abstract}

 The monolithic nature of widely used commodity operating systems means that
 vulnerabilities in one software component potentially compromise the entire
 kernel. Formally verifying these systems, or redesigning them altogether as
 microkernels, according to the principle of least privilege, requires
 significant effort. Researchers have therefore considered compartmentalization
 techniques that minimize or totally avoid changes to existing systems.
 However, current approaches use techniques such as Memory Protection Keys
 (MPKs), necessitating extensive code analysis to ensure security, or use
 virtualization by instrumenting the kernel with calls to the glue code that
 switches compartments.

 In this work, we present \sys{}, a framework that uses hardware-assisted
 virtualization to securely compartmentalize an existing system with minimal to
 no modifications to its source code. Allowed interactions between compartments
 are defined using an access-control policy and strictly enforced
 using Extended Page Tables. Using special {\em sentry} functions,
 \sys{} is able to check all cross-compartment transitions without trapping
 into the hypervisor. We demonstrate the efficacy of \sys{} on a
 compartmentalized Linux network stack using the igc NIC driver. Experiments
 show the overheads of our approach are negligible at MTU-sized packets when compartment boundaries are carefully established to avoid excessive inter-compartment communication.
\end{abstract}




\settopmatter{printfolios=true}
\maketitle
\pagestyle{plain}

\section{Introduction}\label{sect:introduction}

Many operating systems (OSes) in use today are monolithic in structure.
A software fault or security breach in any kernel component may compromise the
entire system. Researchers have studied ways to improve the safety and security
of systems, using techniques such as software fault
isolation~\cite{wahbe-et-al,xfi,castro-sosp-09,mao-sosp-11}, memory-safe
languages~\cite{spin-modula-3,rust}, hardware memory
tagging~\cite{burroughs-b5000,HAKC,ARM-MTE-whitepaper},
capabilities~\cite{intel-ia432,CHERI}, and formal verification~\cite{sel4}.
Others have considered system designs that adhere to the principle of least
privilege, ensuring a software component only has the capabilities to do what
is necessary but no more~\cite{hydra,Cambridge-CAP,levy}.

Micro-kernels~\cite{liedtke-l4,sel4} isolate software components into
domains of the least privilege necessary for their correct operation.
However, the question remains how to achieve the same level of component isolation in
the monolithic systems currently in widespread use. Is it even possible to securely isolate regions within an existing system without significant re-engineering and impact on overall performance? 

Compartmentalization is one approach that is gaining widespread consideration for increased security
and safety of existing monolithic systems~\cite{HAKC,analyse_polp}. Compartmentalization provides
protection domains for separate software components, limiting
access to only the code and data that is necessary for a specific unit of
functionality. Any compartment that makes non-local procedure calls does so
through secure inter-compartment communication mechanisms.

Prior work has considered memory protection
keys~\cite{intel_sdm,bulkhead,connor-usenix-security-20,voulimeneas-eurosys-22,schrammel-usenix-security-22}
as a way to implement kernel and user-level compartmentalization. However,
without an extra ring of protection to enforce access rights to compartments,
privileged instructions within the kernel may bypass protection key security.
For this reason, techniques such as binary rewriting have been used to scan the
kernel for security-critical instructions and replace them with either safe
instructions or those which trap into a trusted in-kernel
monitor~\cite{bulkhead}.

Rather than having to scan and replace security-critical instructions, we
propose a compartmentalization technique that relies on machine virtualization.
While virtualization has long been used to consolidate multiple guest OSes onto
the same physical machine~\cite{popek-goldberg,xen,nova}, it also provides a method to employ an
extra ring of protection to an existing system. Microsoft now uses the Hyper-V
hypervisor to provide virtualization-based security (VBS) for Windows
11~\cite{microsoft_vbs}. Others have used virtualization as a way to enforce
guest compartmentalization~\cite{lvd,epti}, but these works limit compartments
to kernel modules (e.g. for device drivers) and leave the core kernel as one
large trusted codebase.

While some researchers argue that virtualization is too expensive to implement
fine-grained kernel compartmentalization~\cite{bulkhead}, we show how it can
be used efficiently on a modern Linux system. In this paper, we
present a framework called \sys{}, which uses hardware-enforced virtualization
to isolate kernel components into fine-grained compartments. We explain how
\sys{} does this using Intel virtualization extensions (VT-x) and redirection
protection (VT-rp), along with other hardware-enforcement features.

We show how \sys{} implements secure and efficient inter-compartment gate
calls, with the assistance of a small hypervisor that hosts a policy
enforcement manager. This manager is initialized at boot-time with the list of
compartments and their access rights to data objects and kernel functions.
Using special {\em sentry} functions set up by the manager, \sys{} is able to
apply compartmentalization to an existing system with minimal to no
modifications of the kernel. The approach incurs low overheads, as a result of
inter-compartment communication and invalid accesses being handled in the
kernel rather than the hypervisor.

{\bf Contributions:} We describe the implementation of \sys{}, which is built
using the Quest-V partitioning hypervisor~\cite{quest_v, quest_v_vee}. We
explain how \sys{}'s kernel compartmentalization is achieved for both a
source-code annotated version of Linux and also an unmodified system binary.
The overheads of \sys{} are demonstrated for varying numbers of
compartments applied to the network stack. Results show less than
0.01\% throughput overhead at MTU-sized packets for up to 32 compartments,
compared to an unmodified monolithic system, when boundaries are carefully chosen to avoid excessive inter-compartment communication.


\section{Background}\label{sect:background}

\begin{figure}[htbp!]
  \centering
      \includegraphics[width=\columnwidth]{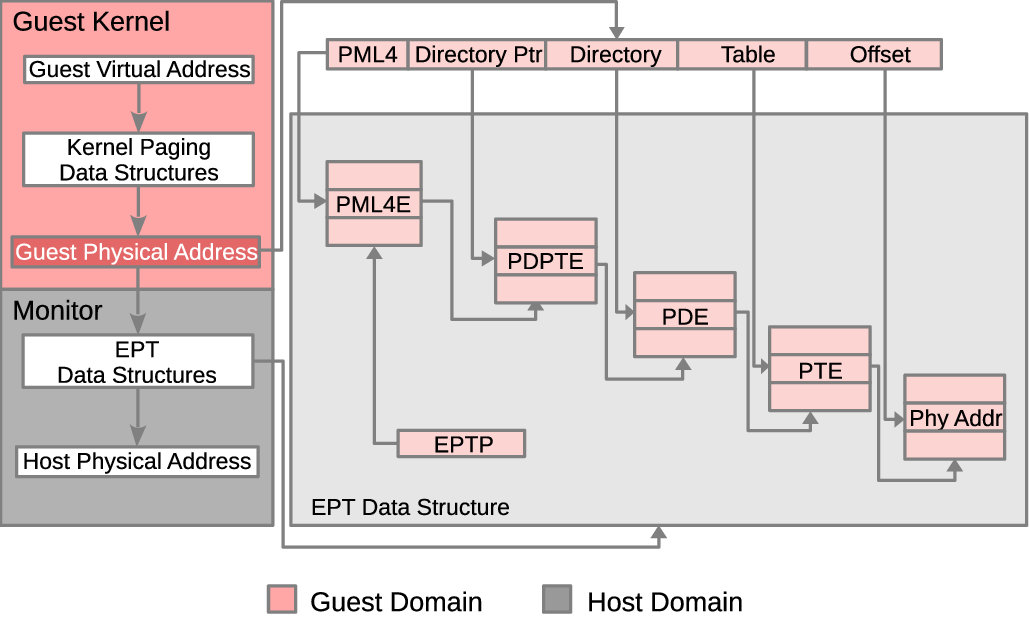}
\caption{Extended Page Table Mechanism}\label{fig:ept}
\Description{Extended Page Table Mechanism in x86 Processors}
\end{figure}

{\bf Virtualization and EPTs:}
Intel VT-x~\cite{intel_sdm} provides hardware support for virtualization,
including Virtual Machine Control Structures (VMCSs) and dedicated VM entry/exit
instructions. One key component is Extended Page Tables (EPTs). Extended Page
Tables, shown in Figure~\ref{fig:ept} are data structures managed by the
hypervisor, which isolate virtual machines using guest physical address (GPA)
to host physical address (HPA) memory mappings. EPTs define both the
permissions and physical memory range accessible to each guest. Each VMCS contains
an Extended Page Table Pointer (EPTP) field that holds the host physical
address of the EPT root table. EPTs are not mapped into, or modifiable by, any guest address
space. Invalid memory accesses trigger traps
into the hypervisor.

While EPTs were originally designed to host multiple
virtual machines on shared hardware, they are also able to isolate
components within a single VM~\cite{epti, lvd, microsoft_vbs}. This approach
moves the root of trust from the kernel to the hypervisor. The hypervisor
creates multiple EPTs for the same VM, with each EPT corresponding to an
isolated compartment. When a compartment executes, the hypervisor switches to
the corresponding EPT, which restricts the accessible instructions and
data.

{\bf Virtual Machine Functions ({\tt vmfunc}):}
Switching EPTs normally requires trapping into the hypervisor, which is
expensive because the hypervisor must save and restore the guest's
architectural state. The Intel {\tt vmfunc} instruction avoids this cost by allowing the guest to
switch EPTs directly. The hypervisor pre-loads an EPTP list with up to 512 EPT
root pointers. The guest issues a {\tt vmfunc} with the desired index to switch the
active EPT while remaining in guest mode. The instruction completes in less
than two hundred cycles. 
The {\tt vmfunc} instruction does not perform any control flow
redirection, which implies that after the EPT switch, execution immediately
continues from the next instruction. Thus, for an attacker's insertion of a
{\tt vmfunc} to succeed, the next instruction has to be mapped into the correct
destination EPT that the attacker intends to switch to. It also does not
perform any control-flow validation. Hence, additional logic is needed to check that a transition is
allowed. Prior work has used an interface definition language to generate this
glue code~\cite{lvd}.

{\bf Virtualization Exceptions:}
EPT violations are typically handled by traps into the hypervisor, but a
subset of them are deliverable to the guest via Virtualization Exceptions
(\#VE). Such exceptions allow the guest to handle certain EPT violations directly,
avoiding the overhead of trapping into the hypervisor. Support for Virtualization Exceptions is activated within the VM-execution control settings of a VMCS. Each EPT
paging-structure entry that references a page additionally uses bit 63 to control where EPT violations are delivered. When bit 63 is set to 1, \#VE delivery is suppressed for the corresponding page and
violations trap to the hypervisor. When cleared, EPT violations are delivered
to the guest as \#VEs. The CPU stores the exit reason, exit qualification
(indicating whether the access was a read, write, or instruction fetch), the
faulting guest linear and physical addresses, and the current EPTP index in a
pre-allocated virtualization-exception information area~\cite{intel_sdm}. We
define a custom \#VE handler that examines this information area, checks
whether the access is allowed, and switches to the appropriate EPT using
{\tt vmfunc}. This avoids trapping into the hypervisor and does not require
generating custom glue code for each compartment transition.

\begin{figure*}[htbp!]
 \centering
     \includegraphics[width=0.9\textwidth]{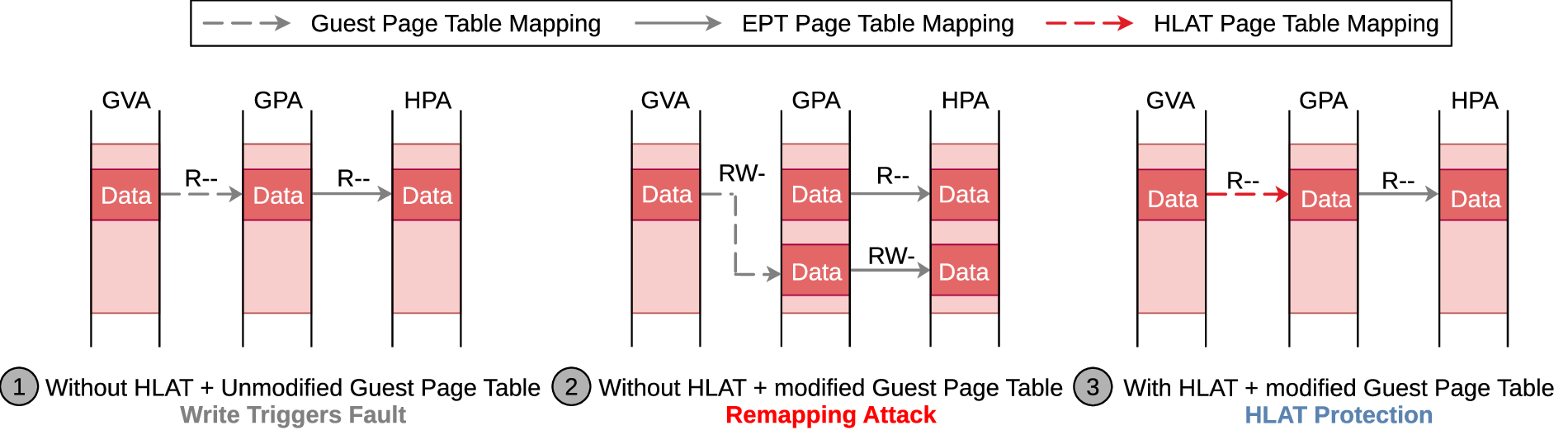}
\caption{Remapping Attack}\label{fig:remap_attack}
\Description{Shows how remapping attack works and how HLAT prevents it}
\end{figure*}

{\bf Hypervisor Managed Linear Address Translation (HLAT):}
EPTs control guest-physical to host-physical translation but cannot control
guest-virtual to guest-physical translation. An attacker who compromises the
guest kernel is able to exploit this gap through a remapping
attack~\cite{satoshi_tanda_2023},
as illustrated in Figure~\ref{fig:remap_attack}. Consider a system
that stores security policy structures, such as access control lists or
compartment permissions, in a page marked read-only in the EPT. Even if the
system is compromised, an attacker cannot modify the original data. However,
the attacker can modify the guest page table to remap the Guest Virtual Address
(GVA) of the sensitive data to a different Guest Physical Address (GPA) that is
marked writable in the EPT. The attacker may then populate this new physical
page with arbitrary policy data, potentially gaining unauthorized access or
escalating privileges.

Intel Hypervisor-managed Linear Address Translation
(HLAT), part of VT-rp (redirection protection), addresses this by allowing the hypervisor to fix a
subset of GVA-to-GPA translations. For pages registered in the HLAT paging
structures, the CPU resolves the translation using the hypervisor-controlled
HLAT entries rather than the guest page tables. Each HLAT entry either
completes the translation or triggers a restart from the guest CR3 paging
structures, so the hypervisor controls only the subset of mappings that require
protection. Guest page table modifications have no effect on HLAT-resolved
translations, which prevents the remapping attack described above. The HLAT
structures themselves are marked read-only in the EPT, so the guest is unable to
modify them.

{\bf Quest-V Partitioning Hypervisor:}
We use Quest-V~\cite{quest_v, quest_v_vee}, to host a compartmentalized version of Linux in this work. Quest-V is a lightweight separation kernel
that statically partitions machine resources among one or more guest sandboxes. Each
sandbox manages its allocated resources, so Quest-V does not cause VM-exits
during normal guest operation. The small hypervisor codebase also reduces the
size of the root of trust.

\section{System Model}\label{sect:system_model}

Our approach follows the mutual distrust model~\cite{sok_compartmentalization},
where no single kernel component is trusted. Instead, the Trusted Computing
Base (TCB) is limited to the hypervisor, with the aim of avoiding privilege
escalation. To that end, the kernel to be compartmentalized should be
statically and dynamically analyzed to determine the compartment
boundaries~\cite{analyse_polp}. A compartment is defined by the following
information:

\begin{itemize}
  \item functions that are executable within the compartment,
  \item functions that can be called in another compartment,
  \item data objects that are readable and/or writeable, and
  \item the execution context, including threads and their stacks, effective user ids ({\tt euid}s) that define capabilities, and the domain (e.g., kernel- versus user-space).
\end{itemize}

Figure~\ref{lst:compartment_def} shows an example of a compartment definition.
While all functions in a compartment can call any other function within the
same compartment, the list of callable functions in \emph{other compartments}
is specified in the \verb|can_call| list. Thus, the \verb|can_call| list
effectively determines the allowed compartment switches. A policy file
containing the definitions of all compartments is given as input to the \sys{}
framework. Ideally, this is provided as part of the boot procedure, before the guest system is loaded and operational. The policy file is then able to be used by the hypervisor to establish guest compartments before the guest itself is fully booted.
Without loss of generality, \sys{} currently establishes compartments in the guest after it has booted, considering boot-time exploits out of scope. 

\begin{figure}[htbp!]
  \begin{lstlisting}[style=Cstyle, basicstyle=\small]
    cmpt_id: 2
    can_execute: func1, func2
    can_read: obj1, obj2, obj3
    can_write: obj1
    can_call: func3
    execution_context: euid = root
  \end{lstlisting} 
  \vspace{-0.2in} 
  \caption{Example Compartment Definition}\label{lst:compartment_def}
  \Description{Example Compartment Definition}
  \vspace{-0.2in}
\end{figure}

\subsection{Threat Model}

We assume an adversary that is able to exploit vulnerabilities anywhere in the
kernel, including both core kernel code and peripheral subsystems, with the
goal of escalating privileges and compromising the entire system.

We adopt a mutual distrust model in which compartments do not place
trust in one another. Each compartment is treated as
potentially compromised and is restricted to accessing only those resources
explicitly permitted by the compartment policy. We assume that the hypervisor,
sentry code, and developer-specified compartment transition policies constitute
our trusted computing base (TCB).

Our threat model makes the following assumptions. First, we assume a secure
boot process that guarantees the kernel image analyzed at design time is
identical to the one loaded at runtime. Second, we focus exclusively on
software-based attacks and do not consider physical exploits such as cold-boot
attacks or hardware tampering.
The scope of this work is limited to static compartment configurations; we do
not currently support dynamically allocated compartments or runtime policy
changes.

\section{Architecture}\label{sect:architecture}

\begin{figure*}[!htbp]
  \centering
  \includegraphics[width=0.65\textwidth]{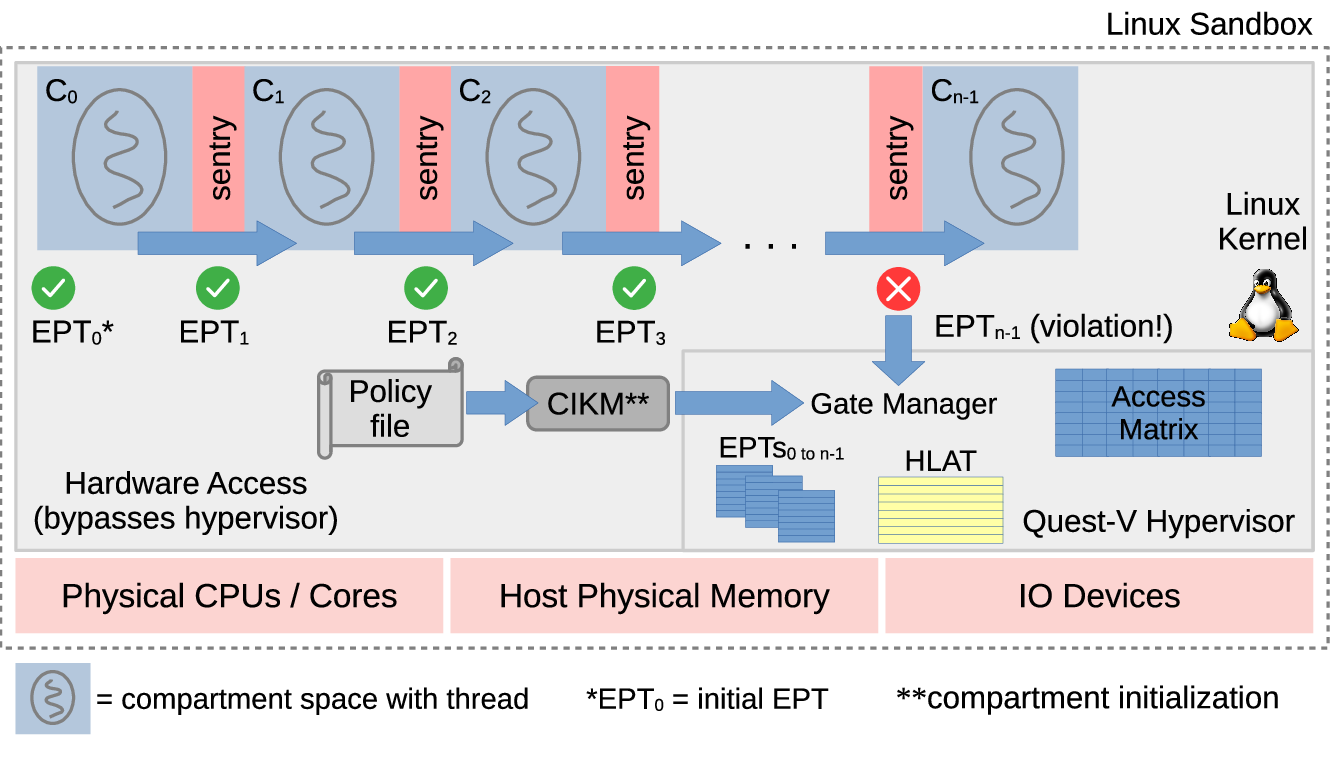}
  \vspace{-0.2in}
\caption{Compartmentalization Architecture}\label{fig:architecture}
\Description{Compartmentalization Architecture}
\end{figure*}


Figure~\ref{fig:architecture} shows the \sys{} compartmentalization framework.
Linux runs as a guest on the Quest-V hypervisor, which partitions hardware
resources at boot-time. As a result, Linux has direct access to its allocated cores,
physical memory, and I/O devices. A lightweight monitor handles VM-exits.
Extended Page Tables (EPTs) control guest-physical to host-physical address
translation. Before booting Linux, Quest-V first creates an initial EPT ($EPT_0$),
where the entire guest physical address space except the monitor itself, is
fully accessible (readable, writable, and executable). This EPT defines the
default compartment $C_0$. A new EPT is then created for each additional
compartment, with restricted mappings.


\subsection{Compartment Initialization}

The policy file, which is given as input to the \sys{} framework, defines
which functions and data objects belong to each compartment, referencing them
by name. EPT construction, however, requires knowledge of the guest physical addresses for each symbol name. A
tamper-proof Compartment Initialization Kernel Module (CIKM), protected by its own EPT mappings, bridges this gap. It parses the
policy file, resolves names to guest virtual addresses,
 and walks
the kernel page tables to obtain the corresponding GPAs. Symbol names to guest virtual addresses are either obtained using \verb|kallsyms_lookup_name|, requiring \texttt{CONFIG\_KALLSYMS}, or by guest code analysis. As knowledge of GPAs is only available after guest page tables are established, we leave secure bootup support for future work. 


The CIKM passes the parsed policy to the hypervisor via a hypercall. The {\em
gate manager}, within the hypervisor, uses this policy to construct a unique EPT for each compartment. It also
populates the EPTP list to enable {\tt vmfunc}-based switching. Access rights for each compartment, having a specific execution context, are recorded in an access matrix.  Once initialization
completes, a thread executing in compartment $C_i$ is restricted to
functions and data objects accessible from within $C_i$.  

The default compartment $C_0$ is handled differently. It uses the initial EPT
($EPT_0$) that is created by Quest-V before Linux boots, to map the
entire guest. When the gate manager creates other compartments,
it modifies $EPT_0$ to unmap data and code that should not be accessible from
$C_0$. Consequently, $C_0$'s policy need not enumerate all accessible
objects—it specifies only objects shared with other compartments and those that
are involved in the transition to another compartment. For example, if functions
in $C_0$ can call \verb|func1| in $C_2$
(Figure~\ref{lst:compartment_def}) and share read access to the \verb|obj2|,
its definition would be as shown in Figure~\ref{lst:C0_def}.

\begin{figure}[!htbp]
  \begin{lstlisting}[style=Cstyle, basicstyle=\small]
    cmpt_id: 0
    can_execute: foo, bar, ...
    can_read: obj2
    can_write: 
    can_call: func1 (cmpt_id=2)
    execution_context: euid = any
  \end{lstlisting}  
  \vspace{-0.2in}
  \caption{Example Default Compartment Definition}\label{lst:C0_def}
  \Description{Example Default Compartment Definition}
\end{figure}

EPTs enforce permissions at 4 KB page granularity. If objects from different
compartments reside on the same page, they receive identical permissions,
potentially causing information leakage.
We currently use compiler directives to enforce page alignment and
discuss sub-page isolation approaches in Section~\ref{sect:discussion}.


\subsection{GVA-to-GPA Translation in \sys{}}

GVA-to-GPA translation uses the kernel page table, which is shared across all
user processes.\footnote{With KPTI enabled, Linux maintains separate kernel
page tables to mitigate Meltdown, incurring TLB flush overhead~\cite{epti}.
EPT-based isolation provides equivalent protection, making KPTI unnecessary.}
For VM-exit-free execution, these paging structures should be accessible to all
compartments. However, the user-space portion of the page table differs per
process. Without HLAT, each process's paging structures would need to be mapped
into every compartment's EPT, requiring EPT updates on the creation of each new
process. HLAT avoids this overhead while also protecting against remapping
attacks. 

\begin{figure}[!htbp]
  \centering
  \includegraphics[width=\columnwidth]{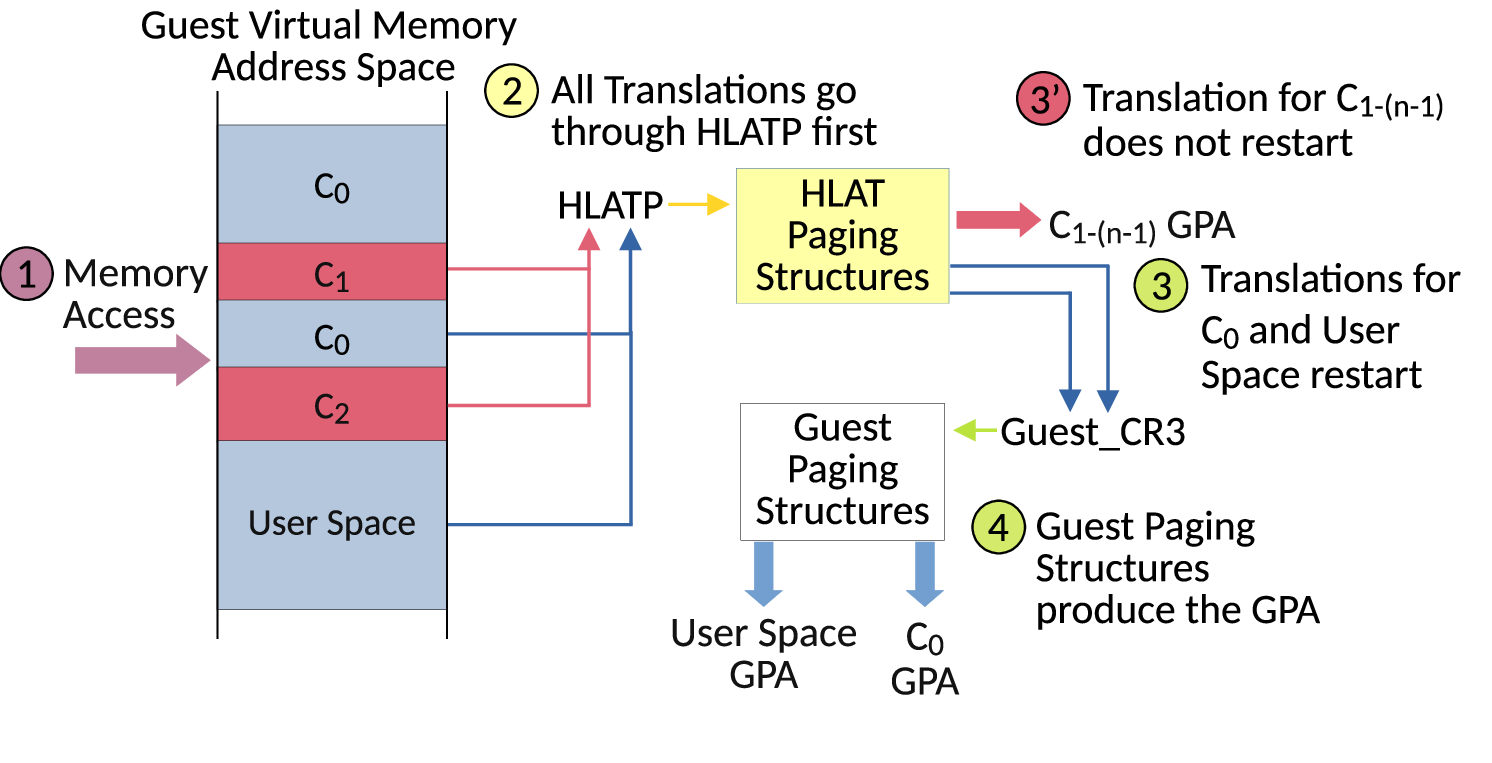}
  \vspace{-0.3in}
\caption{Hypervisor-Managed Address Translation in \sys{}}\label{fig:hlat}
\Description{Flowchart of Hypervisor-Managed Address Translation in \sys{}}
\vspace{-0.1in}
\end{figure}

Figure~\ref{fig:hlat} illustrates \sys{}'s use of HLAT. With HLAT enabled, all
GVA-to-GPA translations begin at the HLAT Pointer (HLATP) rather than Guest
CR3. All memory accesses~\circled{1} start translation at
HLATP~\circled{2}, which references paging structures created by the gate
manager. The CIKM provides GVA-GPA pairs for compartmentalized objects, which
the gate manager uses to populate HLAT entries. Where appropriate, HLAT entries specify that
translation should restart from Guest CR3, allowing \sys{} to include only
compartmentalized objects ($C_1$ to $C_{n-1}$) in the HLAT structure, to
minimize its size. Accesses to these objects obtain their GPA directly from
HLAT~\circled{3'}. Accesses to $C_0$ and user-space memory restart from Guest
CR3, using guest paging structures to produce the GPA~\circled{3},~\circled{4}.

Since the HLAT paging structures have the same number of levels as Guest CR3
and each level contains only the GPA of the next lower-level structure, the
cost of translating a GVA to GPA through HLAT is the same as through Guest
CR3. For objects not mapped into HLAT, the translation restarts from Guest CR3,
adding at most the overhead of walking one additional paging structure if the
translation is not already cached.


EPTs enforce read, write, and execute permissions, but do not specify valid
cross-compartment accesses. Consider \verb|foo| in $C_0$
(Figure~\ref{lst:C0_def}) calling a function \verb|func1| which belongs to
compartment $C_2$. Since \verb|func1| is not mapped in $EPT_0$, the call
triggers an EPT violation. The gate manager inspects the faulting address to
determine if this is a valid cross-compartment call. However, trapping into the
hypervisor costs several hundred to a few thousand cycles
(Section~\ref{subsubsect:ve_overhead}). This overhead is acceptable for invalid
accesses but prohibitive for valid calls, which should incur minimal overhead.

Virtualization Exceptions (\#VE) address this by delivering a subset of EPT
violations directly to the guest rather than the hypervisor. For valid accesses,
{\tt vmfunc} switches the EPT without a hypervisor trap. Beyond switching EPTs,
complete compartment isolation requires changing the thread stack. The \#VE
handler performs both operations, acting as the gate manager's delegate within
the guest. We call this handler the {\em sentry}, which is securely mapped into
every compartment. 

\section{Secure Compartment Switching}\label{sect:sentry}
\subsection{Sentry}

\begin{figure}[htb!]
  \centering
  \includegraphics[width=\linewidth]{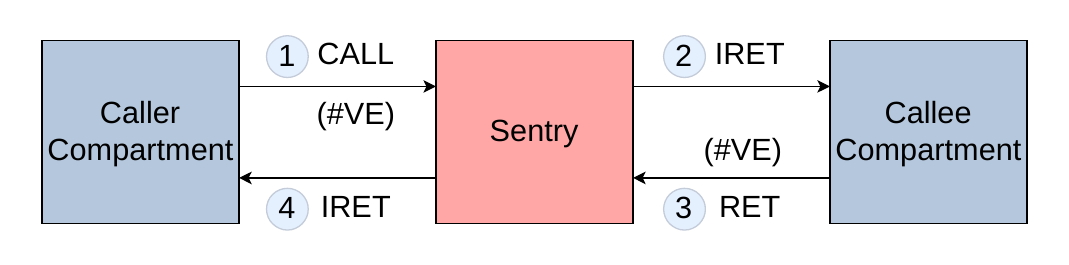}
  \caption{Sentry Call-Return Path}\label{fig:sentry_call_ret_path}
  \Description{Sentry Call-Return Path}
\end{figure}

The sentry handles compartment transitions. On each transition, it verifies
that the access is permitted by the policy, switches the stack and EPT, and
switches control back to the caller when the callee returns.

The flow of control is shown in Figure~\ref{fig:sentry_call_ret_path}. Consider
a call made to a function that belongs to a remote compartment. The hardware
translates the callee function's GVA to GPA via HLAT, then checks the EPT for
the HPA. Since the callee function is not mapped in current compartment, an EPT
violation occurs, invoking the \#VE handler (sentry) (\circled{1} in
Figure~\ref{fig:sentry_call_ret_path}). The sentry saves the caller state,
switches to the callee EPT, pushes the caller return address onto the callee
stack, and sets up an interrupt stack frame such that when it issues an {\tt
iret}, control transfers to the callee function (\circled{2} in
Figure~\ref{fig:sentry_call_ret_path}). When the callee function executes the
{\tt ret} instruction, another EPT violation occurs as the return address on
its stack points to an address in the caller compartment, leading control back
to the sentry (\circled{3} in Figure~\ref{fig:sentry_call_ret_path}). The
sentry restores the caller stack state, and modifies the exception frame that
was pushed in step \circled{1} to again use the {\tt iret} instruction to
return control to the caller (\circled{4} in
Figure~\ref{fig:sentry_call_ret_path}).

Now we explain the working of the sentry in detail, using the pseudocode given
in Procedure~\ref{algo:sentry}. When an unmapped function is called, the
return address is first pushed onto the caller stack, and then as the function
is unmapped a \#VE occurs. The hardware then disables interrupts and pushes
onto the caller stack the following state in the order given: the stack segment
(\verb|SS|), stack pointer (\verb|RSP|, pointing to the caller return address),
flags register (\verb|RFLAGS|), the code segment (\verb|CS|), and the
instruction pointer (\verb|RIP|, pointing to the callee function), before
transferring control to the handler.

\begin{figure*}[htb!]
	\centering
	\includegraphics[width=0.95\textwidth]{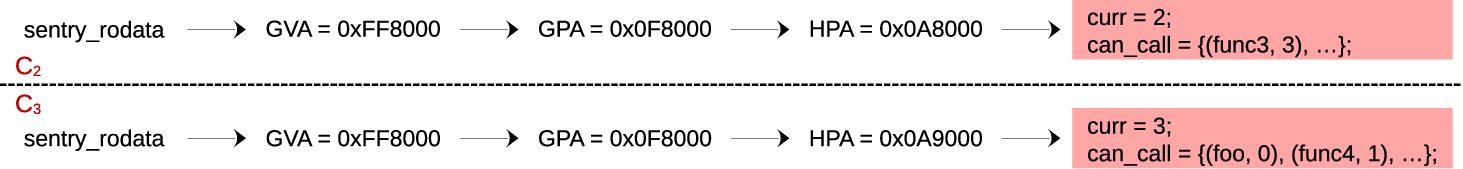}
  \vspace{-0.1in}
	\caption{Example Sentry Read-Only Data Layout}\label{fig:sentry_rodata}
	\Description{Sentry Read-Only Data Layout}
\end{figure*}

Consider the call path first (Lines \ref{line:call_st}-\ref{line:call_end}).
The sentry inspects the \#VE stack frame to determine the callee function
address and the caller return address (Line~\ref{line:call_st}). Then, to
verify a transition, the sentry checks whether the callee address appears in
the caller compartment's \verb|can_call| list (Line~\ref{line:can_call_check}).
Each compartment has a read-only data region (access matrix row) containing its
compartment ID and \verb|can_call| list. This region has the same GVA and GPA
across all compartments but maps to a different HPA as shown in
Figure~\ref{fig:sentry_rodata}. To switch EPTs, the sentry must know the callee
compartment's index. Hence, the \verb|can_call| list is populated by both
the callable function addresses and their corresponding EPT indices,
allowing the sentry to identify the correct callee compartment. In some cases, the same function may map to different compartments
based on execution context. 

In \sys{}, we represent the execution context by the effective user
ID. The current eUID is changed in one of three ways: {\tt execve}
calls, {\tt seteuid} calls, and context switches. {\tt execve} and {\tt
seteuid} are only mapped into $C_0$, so a thread in other compartments cannot
use them to change its eUID. Context switches happen either due to timer interrupts
or explicit invocations of the scheduler, e.g., as a result of a blocking system call.
The scheduler is only mapped into $C_0$, once again preventing changes to the eUID
in other compartments. Additionally, we disable kernel preemption, so timer interrupts that call the scheduler are likewise unable to change the eUID. 

For all non-$C_0$ compartments, we maintain a 1:1 mapping of
execution context to compartment ID. Hence, the sentry only needs to check the
execution context, to determine access rights, when switching from
$C_0$ to another compartment. For this, we use the utilities in the kernel. In
all other cases, the compartment ID implicitly includes the execution context,
and hence the user ID need not be read. The sentry read-only region encodes
these context-to-compartment mappings alongside the \verb|can_call| entries. 

If the callee address is not found in the can\_call list, the access is deemed
invalid and the gate manager is invoked using a {\tt vmcall} to handle the
policy violation (Line~\ref{line:no_match}). If a match is found, the sentry
prepares to switch to the callee compartment. Before switching EPTs, the sentry
pushes any callee-saved registers it will use onto the caller's stack
(Line~\ref{line:save_caller_regs}). According to the System V Application
Binary Interface (ABI) for 64-bit systems, the first six integer arguments
reside in registers, with overflow on the stack. The sentry does not modify
these registers, ensuring that the callee receives correct arguments. We
currently support functions with at most six arguments. Supporting more would
require copying stack-based arguments via a shared page. Pointer arguments must
reference memory shared between caller and callee as specified in the policy. 

\begin{algorithm}[htbp!]
  \caption{Sentry: Call-Return Path}\label{algo:sentry}
  \KwData{\#VE stack frame (SS, RSP, RFLAGS, CS, RIP)}
  \texttt{key} = value at \#VE stack frame RSP\;\label{line:read_call}
  \If{key = $\texttt{MAGIC\_CALL}$ }{ \tcp{Return Path}
    Pop \#VE frame and {\tt MAGIC\_CALL}\;\label{line:pop_callee}
    Clear \#VE mask\;
    Pop caller cmpt\_id\;\label{line:pop_caller_id}
    Save clean callee stack ptr\;\label{line:save_callee_stk}
    {\tt vmfunc} to caller EPT\;\label{line:vmfunc_to_caller}
    Restore caller stack ptr from sentry R/W region\;\label{line:load_caller_stk}
    Pop callee-saved registers\;\label{line:restore_caller_regs}
    Fix \#VE frame instruction and stack pointers\;\label{line:setup_ve_caller}
    {\tt iretq} \tcp*{Return to Caller}\label{line:ret_end}
  }\Else{ 
    \tcp{Call Path}
    Read callee RIP, return addr from \#VE frame\;\label{line:call_st}
    \lIf{curr compt is $C_0$}{read eUID}\label{line:read_euid}
    Look up callee RIP (and eUID) in \texttt{can\_call} list\;\label{line:can_call_check}
    \lIf{no match}{{\tt vmcall} to gate manager}\label{line:no_match}
    Save callee-saved registers on caller stack\;\label{line:save_caller_regs}
    Save caller stack ptr in sentry R/W region\;\label{line:save_caller_stk}
    {\tt vmfunc} to callee EPT\;\label{line:vmfunc_to_callee}
    Load callee stack ptr from sentry R/W region\;\label{line:load_callee_stk}
    Push caller cmpt\_id, \texttt{MAGIC\_CALL} and return addr\;\label{line:push_callee}
    Set up \#VE frame on callee stack\;\label{line:setup_ve_callee}
    Clear \#VE mask\;
    {\tt iretq} \tcp*{Go to Callee}\label{line:call_end}
  }
\end{algorithm}

The sentry then saves the caller stack pointer in the sentry read-write region
(Line~\ref{line:save_caller_stk}). For this purpose, we allocate a
per-compartment stack in the sentry read-write region shared across all
compartments. This data structure is preloaded at initialization with base
pointers for each compartment's stack, allowing the sentry to use identical
code regardless of whether a compartment has prior saved state. While this data
structure is writable in all compartments, corrupting it cannot redirect
control flow: it stores only stack pointers, while the actual stack pages
containing return addresses remain modifiable only within their respective
compartment. It is possible to achieve greater protection, by 
splitting each compartment's read-write region into its own page,
similar to the read-only region setup (Figure~\ref{fig:sentry_rodata}) as each
compartment only needs access to the stack of pointers associated with its own 
stack.

Once the sentry has completed all operations on the caller stack, it uses the
{\tt vmfunc} instruction and the compartment ID obtained from the can\_call
list, to switch to the callee EPT (Line~\ref{line:vmfunc_to_callee}). After the
EPT switch, the sentry loads the callee compartment's stack pointer from the
sentry r/w region (Line~\ref{line:load_callee_stk}). The sentry then sets up a
stack frame to ensure that (1) when it issues an {\tt iret}, control
transfers to the callee function, (2) when the callee issues a {\tt ret} a \#VE
occurs, and (3) the sentry has enough information in the return path to
redirect control to the correct caller address. 

{\bf\noindent Callee Stack Setup:} The sentry pushes the caller compartment ID
onto the callee stack, allowing the sentry to identify the correct caller in
the return path. The return path is handled differently to the call path,
because the sentry does not push a return address onto the stack. To
distinguish between a call and a return, the sentry pushes an 8-byte magic
number, {\tt MAGIC\_CALL}. When the sentry is invoked, it reads the value that
the RSP in the \#VE frame is pointing to (Line~\ref{line:read_call}). The RSP
in the \#VE frame always points to the last value that was pushed onto the
stack before the \#VE occurred. In the case of the call, this is the return
address in the caller. In the case of a return, this is the existing stack
contents. Thus, if this value is equal to the {\tt MAGIC\_CALL}, the sentry
determines that this is a return path, and if not, it treats the value as the
caller return address. This magic number approach ensures that the return
address on the stack is consistent with typical call-return semantics, making
\sys{} compatible with hardware features such as shadow
stacks~\cite{intel_sdm}. Next, the sentry pushes the caller return address onto
the callee stack (Line~\ref{line:push_callee}). Thus, when the callee issues a
\verb|ret|, a \#VE will be triggered, leading control back to the sentry. 

\begin{figure*}[!htb]
  \centering
  \includegraphics[width=\textwidth]{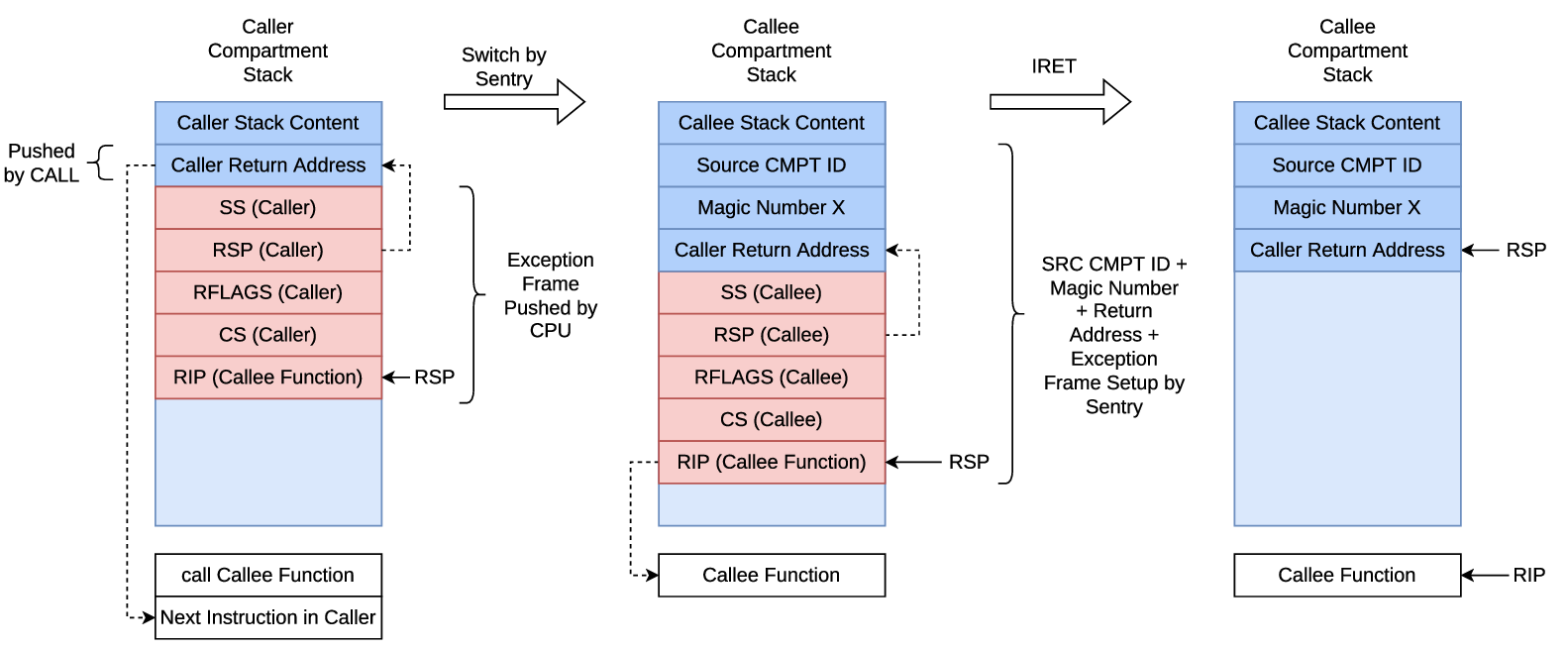}
\caption{Sentry Stack Manipulation: Call Path}\label{fig:stack_call_path}
\Description{Sentry Stack Manipulation in the Call Path}
\end{figure*}

Finally, the sentry sets up a \#VE frame on the callee stack, with the RIP
pointing to the address of the callee function, and the \verb|RSP| pointing to
the caller return address on the callee stack
(Line~\ref{line:setup_ve_callee}). This allows the sentry to redirect control
flow to the callee function using the \verb|iret| instruction. Finally, the
sentry clears the \#VE mask in the Virtualization Exception information area,
re-arming \#VE, before executing the \verb|iret| instruction to "return" to the
callee function (Line~\ref{line:call_end}).
Figure~\ref{fig:stack_call_path} shows the stack manipulations
performed by the sentry in the call path.

When the callee returns, the return address on its stack is popped into RIP.
Since this address points to an instruction in the caller compartment, a \#VE
occurs, invoking the sentry with interrupts disabled and a \#VE frame pushed
onto the callee stack. Once the sentry inspects the magic number, it pops the
\#VE frame and the magic number (Line~\ref{line:pop_callee}) before clearing
the \#VE mask. This ensures that the \#VE handler is invoked at the next EPT
violation. The sentry then pops the caller compartment ID from the callee stack
(Line~\ref{line:pop_caller_id}) for use in the {\tt vmfunc} instruction. Now,
all state information pushed onto the callee stack is fully cleared. This
clean callee stack pointer is saved in the sentry read-write region
(Line~\ref{line:save_callee_stk}). This ensures that, in case of nested
call-returns, we do not run out of stack memory.

\begin{figure*}[!htbp]
  \centering
  \includegraphics[width=\textwidth]{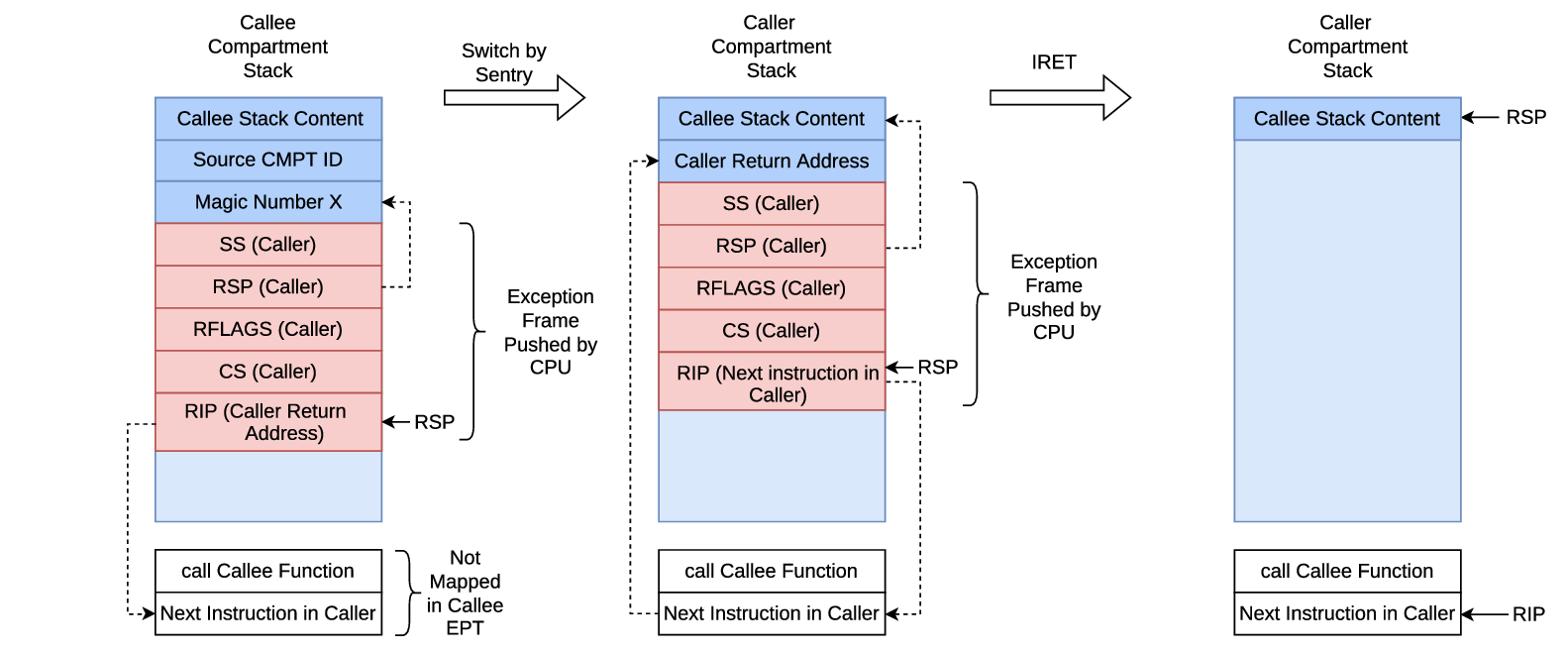}
\caption{Sentry Stack Manipulation: Return Path}\label{fig:stack_ret_path}
\Description{Sentry Stack Manipulation in the Return Path}
\end{figure*}

The sentry then uses the {\tt
vmfunc} instruction to switch back to the caller EPT
(Line~\ref{line:vmfunc_to_caller}). The caller's stack state is restored from
the sentry read-write region (Line~\ref{line:load_caller_stk}). The sentry
restores the callee-saved registers, but preserves \verb|RAX|, which holds the
return value in the 64-bit System V ABI. Finally, the sentry modifies the \#VE
frame that was pushed in the call path to ensure an {\tt iret} diverts control
to the caller. The RIP in the \#VE frame is changed to the return address in
the caller function, and 8 is added to the \verb|RSP| in the \#VE frame
(Line~\ref{line:setup_ve_caller}). Thus, when the sentry issues an \verb|iret|,
the resulting \verb|RSP| skips the caller return address, and control returns
to the correct instruction in the caller. The \verb|iret| instruction
restores the previous state of the \verb|RFLAGS|. Figure~\ref{fig:stack_ret_path} shows the stack manipulations
performed by the sentry in the return path.

While the sentry relies on the callee stack to determine the caller compartment
ID, it must be noted that the underlying EPT mappings ensure that a corrupted
compartment ID by itself cannot result in an access violation. While we
currently do not include \verb|can_return| permissions in the compartment
definition, the design of the sentry lends itself to be easily extended to
support that case, with a check performed in the return path as well as the
call path. The sentry thus handles all compartment transitions without
requiring instrumentation of the compartmentalized code.

\subsection{Interrupt Handling}

\begin{figure}[htbp!]
  \centering
  \vspace{-0.1in}
  \includegraphics[width=\linewidth]{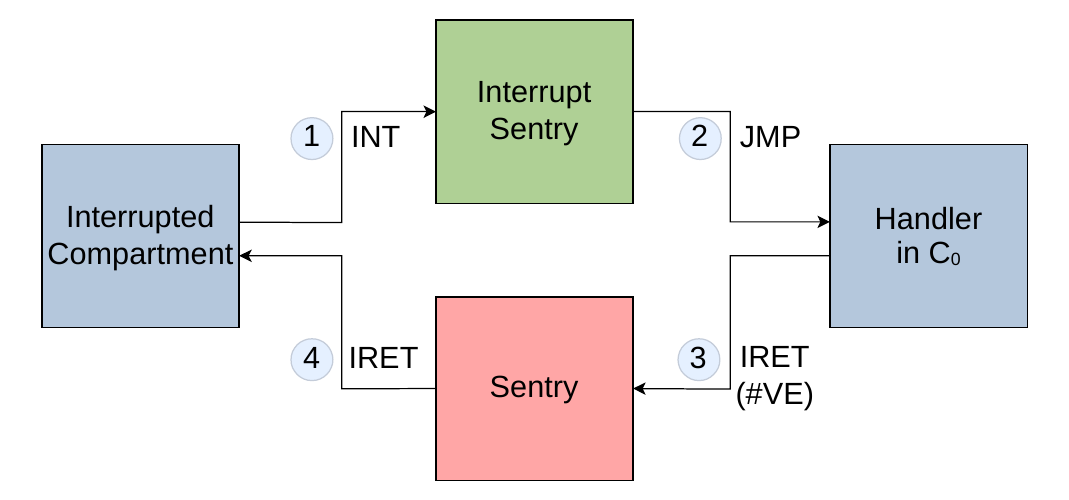}
  \vspace{-0.2in}
  \caption{Interrupt Path in Non-Default Compartments}\label{fig:interrupt_path}
  \vspace{-0.1in}
  \Description{Interrupt path in non-default compartments in \sys{}}
\end{figure}

Interrupts in $C_0$ do not require special handling because $EPT_0$ already
contains all the top half handlers. However, for compartments $C_{1\
to\ (n-1)}$, mapping all interrupt handlers and their dependencies into every
EPT makes it difficult to establish compartment boundaries. The sentry is
capable of disabling interrupts in specific compartments by modifying the
\verb|RFLAGS| in the \#VE frame, but to follow the semantics of the guest, we
support interrupts in all compartments. \sys{} uses special interrupt sentries
to redirect interrupts efficiently to their handlers.
Figure~\ref{fig:interrupt_path} shows the flow of control when an interrupt
occurs in compartments $C_{1\ to\ (n-1)}$.

\begin{figure*}[htbp!]
	\centering
	\includegraphics[width=0.65\textwidth]{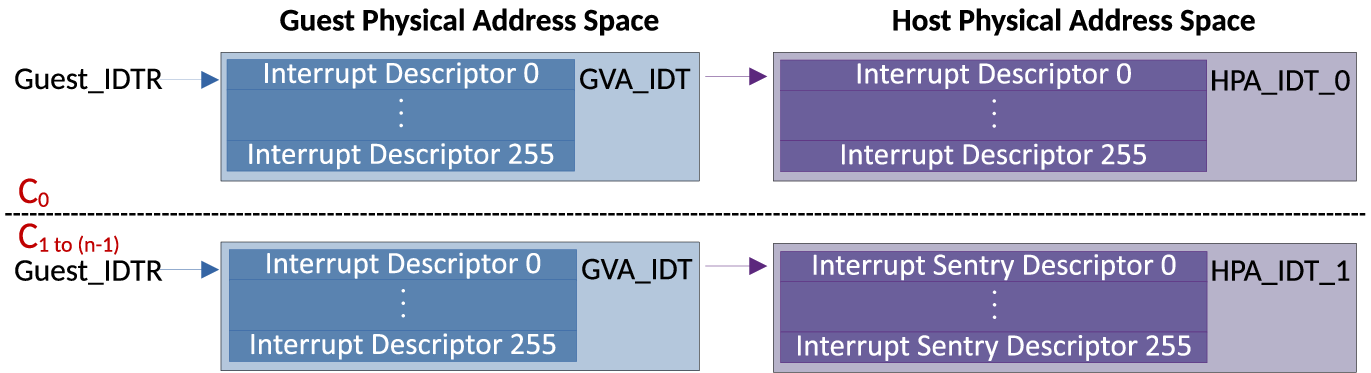}
  \vspace{-0.1in}
	\caption{Interrupt Descriptor Table Setup in \sys{}}\label{fig:idt}
	\Description{Interrupt Descriptor Table Setup in \sys{}}
  \vspace{-0.1in}
\end{figure*}

\begin{algorithm}[htbp!]
  \caption{Interrupt Sentry}\label{alg:int_sentry}
  \KwIn{Interrupt stack frame (SS, RSP, RFLAGS, CS, RIP, optionally Error Code)}
  Push clobbered GP registers onto the current stack\;
  Save stack ptr in sentry R/W region\;
  Switch to interrupt stack\;
  Push interrupted cmpt\_id\;
  Push $\texttt{MAGIC\_INT\_NOERR}$ or $\texttt{MAGIC\_INT\_ERR}$\;
  Set up interrupt stack frame\;
  \If{exception has error code}{Push error code}
  {\tt vmfunc} to $C_0$ (EPT index 0)\;
  {\tt jmp} to appropriate interrupt handler\;
\end{algorithm}

\begin{algorithm}[htbp!]
  \caption{Interrupt Return (Sentry Path)}\label{alg:int_return}
  \KwIn{\#VE frame (SS, RSP, RFLAGS, CS, RIP)}
  \texttt{key} = value at \#VE frame RSP\;
  \If{key = $\texttt{MAGIC\_INT\_NOERR}$ or $\texttt{MAGIC\_INT\_ERR}$}{
  Read source cmpt\_id from interrupt stack\;
  Clear \#VE mask\;
  Save RIP from \#VE frame \tcp{handler may modify it}
  {\tt vmfunc} to interrupted cmpt\_id\;
  Restore stack ptr from sentry R/W region\;
  Update RIP on exception frame to the saved RIP\;
  Pop saved GP registers from the current stack\;
  \If{key = $\texttt{MAGIC\_INT\_ERR}$}
  {$\texttt{RSP} \leftarrow \texttt{RSP} + 8$ \tcp*{Skip error code}}
  {\tt iretq} \tcp*{Return to interrupted code}
  }
\end{algorithm}

The guest virtual and physical addresses of the Interrupt Descriptor Table
(IDT) are left unchanged, but in $EPT_{1\ to\ (n-1)}$ the underlying host
physical page points to \emph{interrupt sentries} instead of the original
handlers, as shown in Figure~\ref{fig:idt}. We allocate a dedicated interrupt
stack page and map it read-writeable in all compartments so that the interrupt
frame is accessible after an EPT switch. Procedure~\ref{alg:int_sentry}
summarizes the working of the interrupt sentry, which handles the interrupt
path, and Procedure~\ref{alg:int_return} shows the interrupt return path, which
is handled by the \#VE handler, i.e., sentry.

When an interrupt occurs in $C_{1\ to\ (n-1)}$, the interrupt sentry is invoked
with interrupts disabled and an interrupt frame pushed onto the current stack.
The interrupt sentry switches to the top of the shared interrupt stack, pushes
the interrupted compartment ID and a magic number (distinct from the one used
for cross-compartment returns), and sets up the interrupt frame, with the
\verb|RIP| pointing to an instruction in the interrupted compartment. After
switching to $C_0$, using a {\tt vmfunc}, the interrupt sentry finally jumps to
the original handler. Since the interrupt sentry always starts from the top of
the interrupt stack, any previous values pushed onto it are overwritten. Hence,
any modifications to the interrupt stack by other compartments will have no
effect.  

\begin{figure*}[!htbp]
  \centering
  \includegraphics[width=\textwidth]{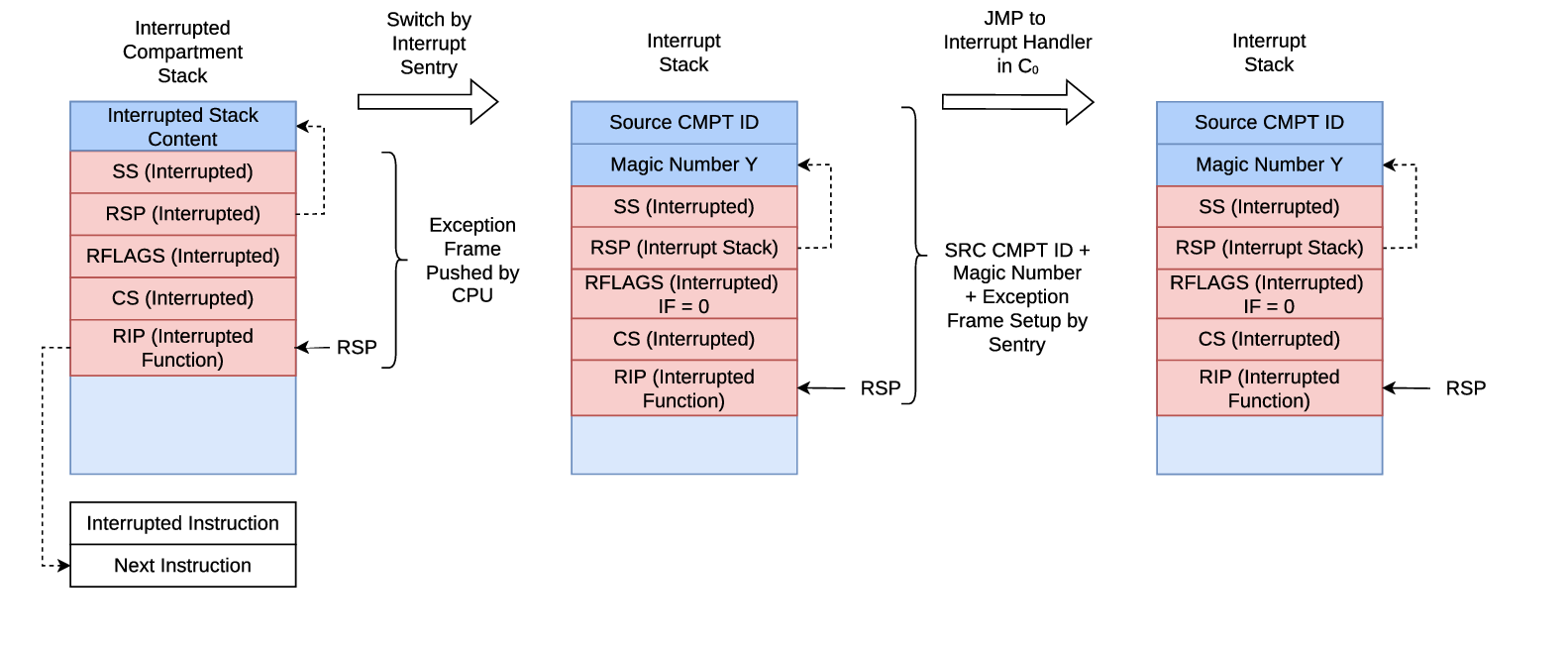}
\caption{Interrupt Sentry Stack Manipulation}\label{fig:stack_int_path}
\Description{Interrupt Sentry Stack Manipulation in the Interrupt Path}
\end{figure*}

\begin{figure*}[!htbp]
  \centering
  \includegraphics[width=\textwidth]{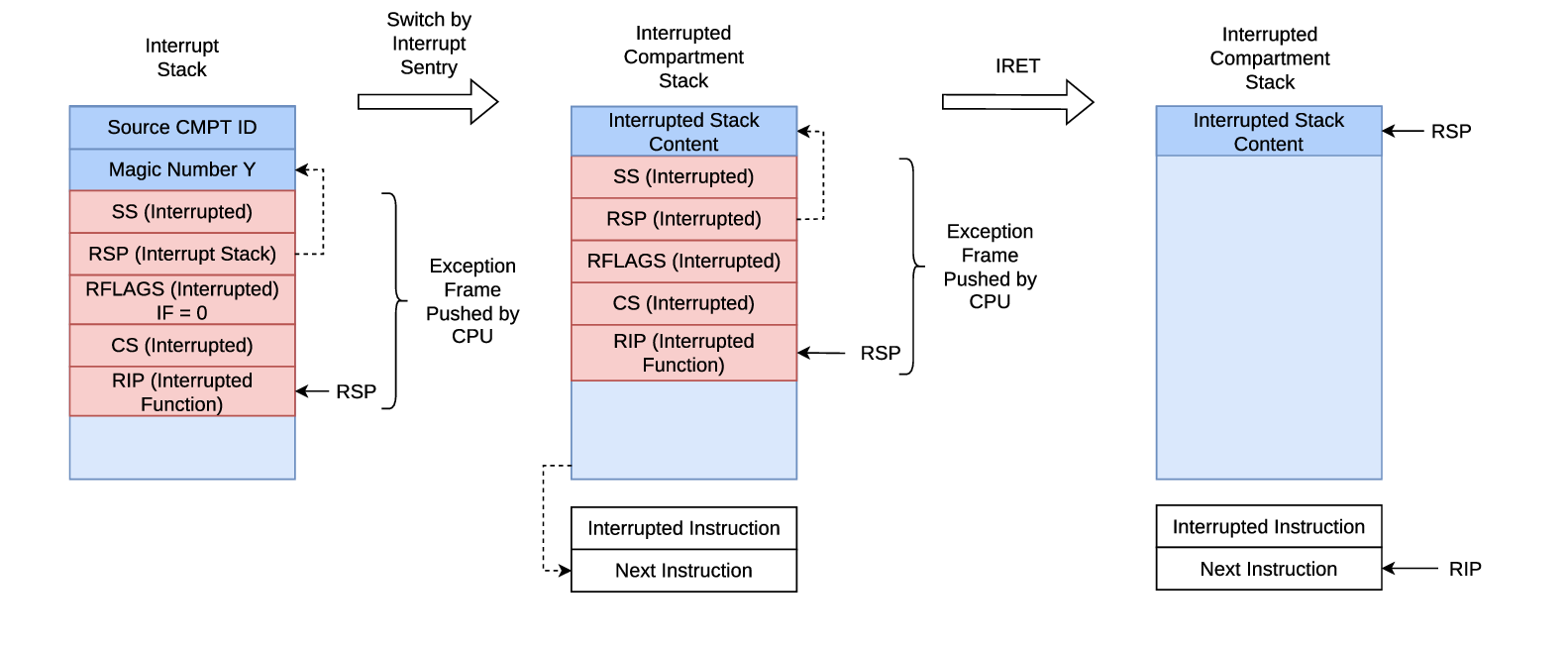}
\caption{Sentry Stack Manipulation: Interrupt Return Path}\label{fig:stack_iret_path}
\Description{Sentry Stack Manipulation in the Interrupt Return Path}
\end{figure*}

When the handler in $C_0$ completes and issues an \verb|iret|, a \#VE is
triggered as the return address is not mapped into $C_0$. This invokes the
sentry with interrupts disabled and a \#VE frame is pushed onto the current
(interrupt) stack. If the value pointed to by the RSP in the \#VE frame matches
the interrupt magic number, the sentry determines that this is an interrupt
return path. It then reads the interrupted compartment ID and uses it to
switch to the right EPT, before issuing an \verb|iret| to resume execution
(Figure~\ref{fig:interrupt_path}). Two magic numbers are chosen from the range
of non-canonical addresses to handle the case where the hardware pushes an
error code, and the case where it does not. Thus, the sentry handles returns
from interrupts in addition to calls and regular returns. 
Figures~\ref{fig:stack_int_path} and~\ref{fig:stack_iret_path} show
the stack manipulations performed by the interrupt sentry in the interrupt path
and the sentry in the {\tt iret} path, respectively.

\section{Experimental Evaluation}\label{sect:exp_eval}

In this section, we evaluate the \sys{} framework. All experiments were
conducted on a Supermicro X13SRN-H motherboard equipped with a 13th Generation
Intel Core i7-1370PE processor. The system runs Quest-V hosting a 64-bit Yocto
Linux Scarthgap guest using the kernel version 6.8.1. To eliminate interference
from concurrent execution and isolate the overhead attributable to our
framework, all experiments were performed with a single active core. First, we
evaluate the overhead of the \sys{} framework and then apply it to a network
driver in the Linux kernel to analyze its impact on the network performance of
the system.

\subsection{Compartment Switch Overhead}
\label{subsec:switch-overhead}

Compartment switch overhead is based on the cost of four
operations: (1)~a Virtualization Exception (\#VE) that traps into the
sentry\label{subsubsect:ve_overhead}, (2)~a {\tt vmfunc} instruction that switches
the EPT, (3)~the sentry logic that checks whether the transition is allowed,
and (4)~the interrupt delivery path that redirects interrupts to the correct
compartment. We measure each operation separately
by executing 10,000 iterations in a kernel loop with interrupts disabled
and record cycle counts with \verb|rdtsc|.
Table~\ref{tab:microbench-summary} reports the median costs, along with the
cost of an undefined opcode exception (\#UD) and an EPT violation trapping into
the hypervisor for comparison. The \#VE is 490~cycles more expensive than \#UD
as a result of the EPT page-walk and Virtualization Exception information area write, but is still
cheaper than a full EPT violation (990~cycles one-way).

\begin{table}
  \centering
  \begin{tabular}{|l|c|c|}
    \hline
    \textbf{Operation} & \textbf{One-way} & \textbf{Round-trip} \\
    \hline
    \#UD (reference) & 448 & 790 \\ \hline
    Virtualization Exception & 938 & 1286 \\ \hline
    EPT Violation (reference) & 990 & 1318 \\ \hline
    {\tt vmfunc} & 142 & - \\ \hline
    Full Sentry Transition & 1782 & 3592 \\ \hline
    Interrupt Handling & 642 & 2958 \\ \hline
  \end{tabular}
  \caption{Median Latency (cycles) of Operations in \sys{}}
  \label{tab:microbench-summary}
  \vspace{-0.3in}
\end{table}

To measure the full sentry cost, we write a kernel module whose device write
function calls another function \verb|test_mod_func| 10,000 times. We then
define two compartments: a test compartment $C_1$ consisting only of the
\verb|test_mod_func|, and the default compartment $C_0$ which consists of the
rest of the kernel and can call \verb|test_mod_func|. The ``Full Sentry
Transition'' row in Table~\ref{tab:microbench-summary} shows the median one-way
(call path) and round-trip (call and return path) cost incurred by the device
write function, excluding \verb|rdtsc| overhead and the execution time of
\verb|test_mod_func| itself. The ``Interrupt Handling'' row shows the one-way
cost of redirecting an interrupt that arrives while the CPU is not in $C_0$
(Procedure~\ref{alg:int_sentry}), and the round-trip cost of interrupt handling
(Figure~\ref{fig:interrupt_path}). To measure the costs of interrupt handling,
we define a custom \#UD handler in $C_0$.

\subsection{Isolated NIC Driver}

To evaluate the overhead of our framework, we partition the kernel
into two compartments: one containing the Intel I225-V NIC driver (igc) and a
default compartment, containing the remainder of the kernel. We choose the igc
driver because it is the NIC present on our test platform and because NIC
drivers produce a high frequency of cross-compartment calls on the data path.
All code in \texttt{drivers/net/ethernet/intel/igc/} constitutes the driver
compartment, while the core kernel, including the networking stack, DMA
subsystem, and memory allocators, constitutes the kernel compartment. All
experiments in this subsection use two Supermicro X13SRN-H machines connected
via Intel I225-V NICs (2.5\,Gbps). One machine is designated as the test
machine running \sys{}-compartmetalized Yocto Linux while the other runs the
same Yocto Linux on bare metal. Since \sys{} does not currently support heap
object isolation, we identify the heap region of the Linux kernel and map it as
read-writeable in both the igc and the default compartments.

First, we analyze the igc data path to determine the number of
cross-compartment switches. Then, we refine the compartment boundary and
amortize the remaining crossings to reduce overhead. Finally, we evaluate the
performance of the unrefined and the refined versions of the igc driver
under the \sys{} framework and compare them against the baseline performance
of the driver without compartmentalization. We disable the driver's periodic
watchdog timer during benchmarking to eliminate infrequent timer-driven
crossings that are irrelevant to data-path performance.


\subsubsection{Kernel Analysis of the igc Data Path}

We statically analyze the igc driver data path in Linux 6.8.1 to identify every
function that crosses the compartment boundary. Table~\ref{tab:igc-exported}
lists all crossing points. Several functions that appear in the driver source
are \texttt{static inline} wrappers defined in kernel headers and do not cross
the boundary. We verify each wrapper to determine whether it ultimately invokes
an exported symbol. For example, \texttt{napi\_gro\_receive()} is
\texttt{static inline} but calls the exported \texttt{gro\_receive\_skb()}.

\begin{table}[t]
  \centering
  \small
  \begin{tabular}{|l|l|l|}
    \hline
    \textbf{Function} & \textbf{Direction} & \textbf{Data Path} \\ \hline
    \texttt{igc\_msix\_ring} & kern $\to$ drv & ISR \\ \hline
    \texttt{napi\_schedule\_prep} & drv $\to$ kern & ISR \\ \hline
    \texttt{\_\_napi\_schedule} & drv $\to$ kern & ISR \\ \hline
    \texttt{igc\_features\_check} & kern $\to$ drv & TX submission \\ \hline
    \texttt{igc\_xmit\_frame} & kern $\to$ drv & TX submission \\ \hline
    \texttt{dma\_map\_page\_attrs} & drv $\to$ kern & TX submission \\ \hline
    \texttt{igc\_poll} & kern $\to$ drv & NAPI poll entry \\ \hline
    \texttt{dma\_unmap\_page\_attrs} & drv $\to$ kern & TX cleanup \\ \hline
    \texttt{napi\_consume\_skb} & drv $\to$ kern & TX cleanup \\ \hline
    \texttt{\_\_memcpy} & drv $\to$ kern & RX \\ \hline
    \texttt{napi\_alloc\_skb} & drv $\to$ kern & RX \\ \hline
    \texttt{eth\_get\_headlen} & drv $\to$ kern & RX ($>$256\,B only) \\ \hline
    \texttt{skb\_add\_rx\_frag} & drv $\to$ kern & RX ($>$256\,B only) \\ \hline
    \texttt{eth\_type\_trans} & drv $\to$ kern & RX \\ \hline
    \texttt{gro\_receive\_skb} & drv $\to$ kern & RX \\ \hline
    \texttt{napi\_complete\_done} & drv $\to$ kern & Poll exit \\ \hline
  \end{tabular}
  \caption{Compartment Switches in the igc Data Path}
  \label{tab:igc-exported}
  \vspace{-0.3in}
\end{table}

Each call to an exported function and its return constitutes two compartment switches. The
igc driver uses NAPI, so ISR and poll entry/exit costs are amortized across a
batch of packets. Transmit (TX) submission incurs 6~switches per packet:
\texttt{igc\_features\_check} (called by the kernel on every outgoing packet
before \texttt{igc\_xmit\_frame}) and \texttt{dma\_map\_page\_attrs} each
contribute a call--return pair in addition to \texttt{igc\_xmit\_frame} itself.
TX cleanup incurs 4~switches per packet. On the receive (RX) path, four exported
functions cross the boundary per packet: \texttt{\_\_memcpy}, \texttt{napi\_alloc\_skb},
\texttt{eth\_type\_trans}, and \texttt{gro\_receive\_skb}. Packets larger than 256\,bytes on the
wire additionally invoke \texttt{eth\_get\_headlen} and
\texttt{skb\_add\_rx\_frag}. Including the amortized poll overhead
(\texttt{napi\_complete\_done}), the per-packet totals are 10~switches for
small RX packets and 14~for large ones
(Table~\ref{tab:igc-switches-opt}, Baseline column).
When the CPU is in the driver compartment, an additional 2~switches are needed
for interrupt delivery (Figure~\ref{fig:interrupt_path}).

\subsubsection{Reducing Compartment Crossings}
\label{subsubsec:igc-opt}

The baseline compartment boundary between the igc driver module and the
rest of the kernel is determined by the module's symbol table: every call
to or from an exported kernel symbol crosses the boundary.  Many of these
crossings are unnecessary---the called function either performs trivial
work that could execute in the caller compartment, or is a
platform-specific no-op.  Each call-return pair nonetheless incurs two
compartment switches ($\sim$3592~cycles round-trip, Table~\ref{tab:microbench-summary}).

We reduce crossings by applying two strategies. First, we \emph{refine the
compartment boundary} by relocating functions across it: small, stateless
kernel functions are moved into the driver compartment, while driver entry
points that perform only kernel-relevant work are replaced with kernel-side
stubs.  Second, we \emph{amortize} the remaining crossings by batching
per-packet kernel calls into per-poll helpers that cross the boundary once per
batch. It is possible to implement both refinement strategies by modifying the
underlying EPT mappings to refer to replacement code in host memory. However,
our prototype approach modifies the Linux source code directly, for simplicity.

{\bf Boundary Refinement:}
We relocate functions across the compartment boundary in both directions.

\emph{Into the driver:}
On x86 with direct DMA and no IOMMU, \texttt{dma\_map\_page\_attrs} reduces to
a single arithmetic operation that returns the physical address as the DMA
address, and \texttt{dma\_unmap\_page\_attrs} is a complete no-op; we compile
the mapping logic into the driver and remove the unmapping call, eliminating
2~switches per TX submission and 2~per TX cleanup.  On the RX path, \texttt{skb\_add\_rx\_frag}
(shared-heap only), \texttt{eth\_type\_trans} (pure header parsing),
and the kernel's \texttt{\_\_memcpy} are each moved into the driver compartment.
\texttt{eth\_get\_headlen} depends on the flow dissector
\texttt{\_\_skb\_flow\_dissect}; we copy the dissector into the driver
after removing unused BPF code paths,
mapping one read-only kernel global into the driver EPT.  These
relocations eliminate the size-dependent RX crossing split: after
refinement, the RX path incurs the same crossing count regardless of packet
size.  The driver's TCB grows, but all relocated code is stateless packet
processing or arithmetic on shared-heap data; the driver already has access to
page structures through the shared heap, so the driver gains no new capability.

\emph{Out of the driver:}
The top-half interrupt handler performs only an MMIO write to the NIC's
interrupt throttle register and a call to schedule NAPI---trivial work wrapped
in 6~handler gate crossings costing $\sim$10,776~cycles.  We replace it with a
kernel-side trampoline that performs both operations without entering the driver
compartment, eliminating all 6~handler crossings.  The interrupt delivery
overhead (2~switches when the CPU is in the driver compartment;
Figure~\ref{fig:interrupt_path}) remains, as it is part of our framework's
interrupt handling mechanism.  Similarly, the kernel calls
\texttt{igc\_features\_check} on every outgoing packet to validate offload
features---pure skb header inspection with no driver-private state.  We replace
it with a kernel-side stub, eliminating 2~switches per TX packet.  The security
trade-off is that the kernel compartment gains write access to the NIC's ITR
register, which controls only interrupt coalescing timing, and read access to
skb headers that it already constructed.

{\bf Batch amortization:}
Three per-packet kernel calls remain after boundary refinement:
\texttt{napi\_alloc\_skb} (RX allocation), \texttt{napi\_consume\_skb} (TX
cleanup), and \texttt{napi\_gro\_receive} (RX delivery to the protocol stack).
We replace each with a kernel-side batch helper.  A pre-allocation helper
allocates an entire poll budget's worth of \texttt{sk\_buff}s in a single
crossing at the start of each NAPI poll; a second helper frees any unused
remainder.  For TX cleanup, we collect completed \texttt{sk\_buff} pointers
during the cleanup loop and pass them to a batch helper in one crossing,
amortizing the per-packet cost from 2~switches to $\sim$0.02~switches
(2~switches per 128~packets).  For RX delivery, we defer
\texttt{napi\_gro\_receive} calls, collecting completed \texttt{sk\_buff}s
during the receive loop and delivering the entire batch in a single crossing
at the end of the poll---amortizing from 2~switches per packet to
$\sim$0.03~switches per packet (2~switches per budget of 64~packets).  The deferral is
safe: the batch helpers perform the same operations as the individual calls,
touch no descriptor ring state, and the driver does not access an
\texttt{sk\_buff} after handing it to the helpers.  We also skip the RX batch
allocation during TX-only polls by checking the first RX descriptor before
crossing. In total, the amortized RX path incurs $\sim$0.06 switches per packet.

Table~\ref{tab:igc-switches-opt} shows the compartment switch counts
before and after applying boundary refinement and batch amortization.

\begin{table}[htbp!]
  \centering
  \begin{tabular}{|l|l|c|c|}
    \hline
    \textbf{Path} & \textbf{Frequency} & \textbf{Baseline} & \textbf{Reduced} \\ \hline
    ISR & per interrupt & 6 & 0 \\ \hline
    TX submission & per packet & 6 & 2 \\ \hline
    TX cleanup & per packet & 4 & $\sim$0.02 \\ \hline
    RX ($\leq$256\,B) & per packet & 10 & $\sim$0.06 \\ \hline
    RX ($>$256\,B) & per packet & 14 & $\sim$0.06 \\ \hline
    Poll entry/exit & per poll & 2--4 & 2--4 \\ \hline
  \end{tabular}
  \caption{Compartment Switch Counts Before and After Crossing Reduction}
  \label{tab:igc-switches-opt}
  \vspace{-0.3in}
\end{table}

After boundary refinement and batch amortization, the only remaining per-packet
crossing is \texttt{igc\_xmit\_frame} entry/return (2~switches per TX packet);
all RX per-packet crossings are fully amortized into per-poll batches.

\subsubsection{Experiment 1: Single-Flow Throughput}

To understand the compartmentalization overhead on the igc driver data path, we
perform the netperf~\cite{netperf} {\tt UDP\_STREAM} test for 30 seconds at
different message sizes. We test three configurations of the test machine:
(1)~a \emph{baseline} kernel without compartmentalization running on Quest-V,
(2)~an \emph{unrefined} compartmentalized kernel with all original boundary
crossings, and (3)~a \emph{refined} compartmentalized kernel with the boundary
refinement and batch amortization described in Section~\ref{subsubsec:igc-opt}.
For TX, the netperf client runs on the test machine and the server on an
unmodified machine. For RX, the client runs on the unmodified machine and the
test machine acts as the server. We disable interrupts in
the igc compartment using the sentry to minimize compartment crossings in both
compartmentalized configurations. We vary the netperf message size
(\texttt{-m}) from 64\,bytes to 1472\,bytes (the Maximum Transmission Unit, MTU),
spanning six sizes: 64, 128, 256, 512, 1024, and 1472.

\begin{figure}[t]
  \centering
  \includegraphics[width=\columnwidth]{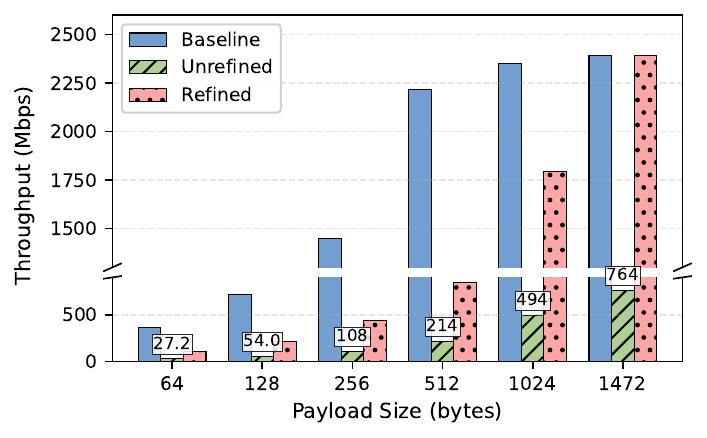}
  \vspace{-0.3in}
  \caption{Single-Flow TX Throughput}
  \label{fig:tx-comparison}
  \Description{Single-Flow TX Throughput (Mbps) vs.\ Payload Size bar graph}
\end{figure}

{\bf TX throughput:}
Figure~\ref{fig:tx-comparison} shows TX throughput for all three
configurations.  When the payload is 1472\,bytes, the 
refined compartmentalized configuration reaches the line rate ($\sim$2393\,Mbps), matching the
baseline, while the unrefined configuration only achieves
764\,Mbps (almost 70\% reduction).  At 64\,bytes, where per-packet overhead
dominates, the unrefined configuration delivers only 27\,Mbps (vs.\ 359\,Mbps
baseline, a 92\% reduction). In comparison, the refined configuration reaches 108\,Mbps
(70\% reduction). The remaining TX overhead is dominated by the two irreducible
\texttt{igc\_xmit\_frame} switches per packet
(Table~\ref{tab:igc-switches-opt}).

\begin{figure}[t]
  \centering
  \includegraphics[width=\columnwidth]{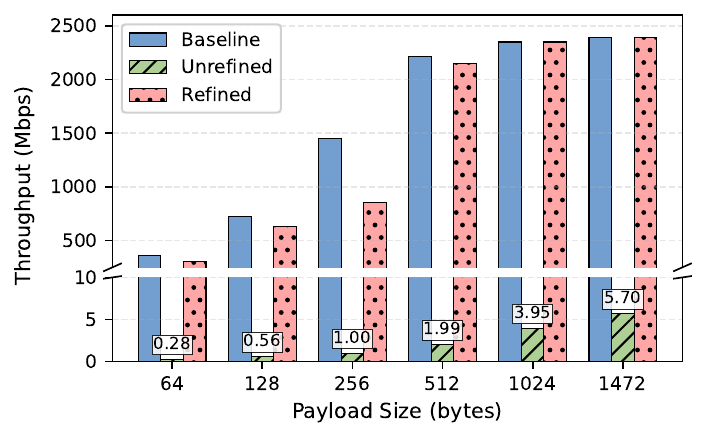}
  \vspace{-0.3in}
  \caption{Single-Flow RX Throughput}
  \label{fig:rx-comparison}
  \vspace{-0.05in}
  \Description{Single-Flow RX Throughput (Mbps) vs.\ Payload Size bar graph}
\end{figure}

{\bf RX throughput:}
Figure~\ref{fig:rx-comparison} shows the RX throughput. The unrefined configuration
delivers $<$6\,Mbps at all packet sizes, due to the large number of compartment
switches in the typical RX path. The refined configuration reaches 304\,Mbps at
64\,bytes (84\% of the 360\,Mbps baseline) and matches the line rate from
512\,bytes onward. The RX throughput exceeds TX throughput at small packet sizes because batch
amortization reduces the number of compartment switches in the reception path to $\sim$0.06~switches/packet
(Table~\ref{tab:igc-switches-opt}), while the transmission path retains two irreducible per-packet
switches. The RX throughput dip at 256\,bytes in the refined
configuration corresponds to the \texttt{IGC\_RX\_HDR\_LEN} threshold, where
the driver switches from a linear copy to a page-fragment mode with flow
dissection. The additional per-packet cost of this path is negligible in
the baseline, but accounts for a $\sim$40\% throughput reduction under
compartmentalization at that packet size.  Larger payloads amortize this
cost.

\subsubsection{Experiment 2: Multi-Compartment Evaluation}

Next, we measure how throughput changes with the number of active compartments,
by setting up a multi-threaded TX test. Each igc compartment shares the
same driver code and heap; only the per-compartment stack and execution
context, in this case the thread ID, differ. This experiment evaluates the cost
of the sentry policy lookup and any TLB or cache pressure from maintaining
multiple EPT contexts. We use the \emph{refined} compartmentalized
configuration (Section~\ref{subsubsec:igc-opt}), which incurs 2~switches per TX
submission. We launch $N$ concurrent netperf {\tt UDP\_STREAM} instances on the
test machine, each assigned to a distinct compartment, with a per-packet
payload size of 1472\,bytes. All instances send to the same port of the
same unmodified receiver. A baseline system without
compartmentalization is run with the same number of $N$ threads, for comparison.

We only evaluate packet transmission as it runs in a process context with each
\texttt{sendto} carrying the calling thread's compartment identity. Hence, $N$
concurrent senders present $N$ distinct identities to the sentry.  Packet
reception, in contrast, runs in NAPI softirq context---the driver's
\texttt{igc\_poll} dequeues all packets from a shared receive ring regardless
of destination flow, and demultiplexing by port occurs in the kernel's
transport layer above the compartment boundary. Additional compartments
do not affect the RX data path.

\begin{table}[t]
  \centering
  \small
  \begin{tabular}{|c|r|r|}
    \hline
    \textbf{$N$} & \textbf{Baseline (Mbps)} & \textbf{Compartmentalized (Mbps)} \\ \hline
    2--32 & 2392.77--2395.78 & 2392.67--2394.55 \\ \hline
  \end{tabular}
  \caption{Multi-Compartment TX Aggregate Throughput}
  \label{tab:igc-multicmpt}
  \vspace{-0.3in}
\end{table}

Table~\ref{tab:igc-multicmpt} shows that the aggregate throughput remains at
line rate ($\sim$2393\,Mbps) for all tested values of $N$, with less than
2\,Mbps difference between the baseline and the compartmentalized
configuration. All threads achieve approximately the same throughput in all
cases. Increasing the number of igc compartments from~2 to~32 introduces no
measurable aggregate throughput difference, consistent with no TLB or cache
contention from maintaining multiple EPT contexts at this packet size.


\section{Discussion}\label{sect:discussion}

{\bf Resilience to Attacks:}
There is nothing to stop an attacker from inserting arbitrary {\tt vmfunc}
calls into an instruction sequence. However, as they do not divert control flow
by themselves, the instruction immediately after a {\tt vmfunc} will trigger an
EPT violation, unless it is mapped into the target compartment. Control-Flow
Enforcement Technology (CET) prevents an attacker from jumping to an arbitary
sentry address, to avoid the {\tt can\_call} check, or corrupt the return
address in the callee stack. With CET, legitimate indirect calls or jumps are
followed by the {\tt endbr} instruction. The technology also uses shadow stacks
to verify valid return addresses.

{\bf Compartment-Aware Kernel Drivers:} 
While drivers are a prominent source of vulnerabilities in
Linux~\cite{linux_vulnerabilities}, our analysis shows that the design of the
igc NIC driver does not easily lend itself to compartmentalization, due to significant interaction with core kernel code. Other complications relate to the handling of RX and TX cleanup using the same softirq handler, making it difficult to (a) provide differentiated quality-of-service~\cite{usb_diff_services}, and (b) isolate the handling of received data for different users.
Our approach to refine compartment boundaries requires altering the logic
within separate compartments using interposed code that is mapped by EPTs to
the original guest physical addresses. Similar interposition techniques are
able to improve service quality and isolation between users. 

{\bf Policy Generation:}
The granularity of compartments needs to be determined by carefully considering
the trade-off between security and cost of enforcement. There is little utility
in splitting each function into its own compartment if the resultant overhead
is prohibitive. Another factor that should be considered when determing
compartment size is the amount of global data shared between compartments.
Shared data should be minimized to ensure a small attack surface. Any pointers
passed across compartment boundaries should preferably be sized such that their
values fit within registers exchanged between caller and callee.

{\bf Sub-Page Isolation:}
EPTs enforce permissions at page granularity. Such permissions fail when functions or static data
from different compartments share a physical page.
We currently use compiler directives applied to guest kernel source code to page-align
compartmentalized objects (Section~\ref{sect:architecture}). Alternatively, source code annotations are unnecessary if the hypervisor allocates a separate copy of each shared page
per compartment, containing only that compartment's code and data at their original
offsets. The remaining addresses are filled with trapping values (for code) or
zeroes (for data). This allows sub-page isolation without guest kernel
modification.

{\bf Compartment Switching Mechanism:}
\sys{} uses Virtualization Exceptions to facilitate compartment
switches without source code or binary modification. However, source code
annotations or binary rewriting has the potential to reduce the overhead of
compartment switching. 

{\bf Implicit vs. Explicit Gate Calls:}
\sys{} uses a sentry to implement gate calls that mediate transitions between
compartments. We consider this approach an implicit gate calling mechanism
because it does not require source code or binary modification of the guest.
An explicit gate call would require guest code annotations to jump to a
gate manager that validates compartment switches. One may wonder why it makes
sense to support implicit gate calls when the establishment of compartments and
policies requires source code analysis. We believe implicit gate calls are
beneficial when the policy generation and enforcement are handled by
separate entities. Moreover, policy updates are easily handled with an
implicit gate calling mechanism, while they would require potential source code
changes with an explicit approach.

{\bf Heap Object Isolation:}
All kernel code shares a common heap. As stated above, EPT permissions apply only at page
granularity, so heap objects from different compartments that share a page
cannot easily be isolated using only EPTs. TME-MK~\cite{intel_tme_mk} provides
sub-page read isolation by assigning a different key ID per compartment in each
EPT. The encryption engine operates at cache-line granularity, so accesses with
an incorrect key ID return garbled data. TDX~\cite{intel_tdx} adds write
protection using Message Authentication Code (MAC) verification. However, TME-MK cannot protect objects
smaller than a cache line (64\,bytes), and TDX's behavior on a MAC mismatch
varies across platforms~\cite{tme_box}. An alternative approach that does not require such esoteric hardware features is to implement per-compartment heap-memory allocation.
Each compartment is then assigned a dedicated heap region unmapped from other compartments' EPTs. All original dynamic memory allocation requests are then redirected to compartment-specific allocators using the same interposition techniques described earlier. This avoids kernel source code or binary modifications.

{\bf Multi-core Support:}
We enabled only one core to avoid interference from impacting the experimental
results, but \sys{} is runnable on a multi-core system as long as the tasks in
non-default compartments remain on the cores where they began execution. The
VMCS tracks the current EPT on a per-core basis. Hence, to support task
migration, the hypervisor needs to create the same EPTP list on all cores.
In addition to calls, returns and interrupt returns, the sentry should also
handle \#VEs that occur when a task migrates to a core where the current EPT
does not map its data or code.

\vspace{-0.1in}
\section{Related Work}\label{sect:rel_work}

Lefeuvre et al.~\cite{sok_compartmentalization} classify compartmentalization
approaches into sandbox, safebox, and mutual\st{-} distrust models. Lim et
al.~\cite{compartment_survey} survey kernel compartmentalization techniques
along similar lines. Our work follows the mutual\st{-} distrust model, where all
kernel components are treated as equally untrusted. \sys{} differs from prior kernel compartmentalization systems in that
compartment boundaries are enforced from outside the guest, at a privilege level
no guest code can subvert. The guest kernel executes unmodified and is unaware
of compartmentalization; boundaries are defined by the hypervisor's EPT
configuration and a generic sentry that handles all compartment transitions
without per-compartment glue code. Roessler et al.~\cite{analyse_polp} present
$\mu$SCOPE, a methodology for analyzing least-privilege compartmentalization in
large codebases. Our work is compatible with such a mechanism that analyzes a
monolithic kernel to determine the right compartment boundaries.

{\bf Virtualization-based Approaches:}
EPTI~\cite{epti} uses EPT switching as an alternative to kernel page table
isolation for Meltdown mitigation but does not provide intra-kernel
compartmentalization. LXDs~\cite{lxds} isolate kernel subsystems into
lightweight execution domains. Per-module glue code is generated from an
interface definition language (IDL) to handle data marshalling across
compartment boundaries. LVD~\cite{lvd} builds upon LXDs by using {\tt vmfunc} to make
compartment transitions more efficient. KSplit~\cite{ksplit} automates the
generation of this IDL-based marshalling code for device drivers. All three
systems require custom glue code compiled into the kernel for each compartment
boundary, and compartmentalization is limited to kernel drivers.

In contrast, \sys{} uses a combination of Virtualization Exceptions and {\tt
vmfunc} to enable compartment switches without source code or binary
modification. These features are used to implement a sentry mechanism that
interposes on function calls that cross compartment boundaries. \sys{} uses
HLAT to ensure that the guest does not modify its own page tables in a way that
violates the system security policy, avoiding the need for glue code to achieve
the same goal. The sentry logic maintains a centralized access control matrix
that ensures all compartment crossings adhere to legitimate execution paths. 

Microsoft's Virtualization-Based
Security~\cite{microsoft_vbs} uses Hyper-V to protect sensitive Windows
components but does not support mutual distrust between arbitrary kernel
components. xMP~\cite{xMP} provides memory protection primitives
within a VM, allowing guests to isolate sensitive data in disjoint domains using
EPT subspaces. While xMP also uses Virtualization Exceptions and {\tt vmfunc}, it
targets individual data structures (e.g., page tables, credentials) rather than
general compartmentalization, and does not provide a sentry mechanism for
compartment transitions with access control.

{\bf Other Isolation Mechanisms:}
ERIM~\cite{erim} uses Intel Memory Protection Keys (MPK) for in-process
isolation on x86. BULKHEAD~\cite{bulkhead} applies Protection Keys for Supervisor (PKS) access to kernel compartmentalization, but requires binary rewriting to neutralize \verb|wrpkru| and other instructions that could bypass protection
keys. Connor et al.~\cite{connor-usenix-security-20} and Voulimeneas et
al.~\cite{voulimeneas-eurosys-22} use MPK for user-space sandboxing. Schrammel
et al.~\cite{schrammel-usenix-security-22} use memory tagging with
cryptographic integrity on commodity x86. PKS and MPK operate at the kernel
privilege level, so a compromised kernel can execute privileged
instructions to bypass protection enforcement. All of these approaches therefore require
binary analysis or rewriting to neutralize such instructions. In \sys{},
enforcement is in the hypervisor: no guest instruction is able to modify EPT
permissions regardless of privilege level, so no instruction scanning or binary
rewriting is needed.
EKC~\cite{ekc} provides a portable kernel compartment that enforces isolation
but, like other methods that do not use a higher privilege level, requires
code analysis to eliminate instructions that subvert the protection mechanism.

In other work, Mondrix~\cite{mondrix} applies Mondriaan memory protection to the Linux kernel.
SVA~\cite{secure_virt_arch} defines a virtual instruction set that enforces
memory safety and control-flow integrity for system code.

\vspace{-0.1in}
\section{Conclusion and Future Work}\label{sect:conclusion}

This work presents \sys{}, a framework that uses hardware-assisted
virtualization to compartmentalize a monolithic kernel. By enforcing isolation
through EPTs, \sys{} is able to compartmentalize a monolithic guest without
modifying its source code or binary image.
\sys{} uses a generic
sentry mechanism that handles all compartment transitions in the guest using
Virtualization Exceptions and EPTs. Sentry execution avoid traps into a hypervisor-based gate manager, and allows direct access to guest state. This avoids expensive guest page table walks to perform compartment switching, if it were performed within the hypervisor. However, Virtualization Exceptions are shown to incur more overhead than other exceptions, as the CPU must walk the
EPT table and write to a special information area the reason for the exception. This is an area where future hardware optimizations would be beneficial to virtualization-based compartmentalization approaches such as \sys{}.

\sys{} is evaluated on a compartmentalized Linux network stack using the igc
NIC driver. With up to 32 separate execution contexts, results show that compartmentalization incurs negligible overhead for MTU-sized packets. By analyzing compartment crossings and redrawing boundaries,
\sys{} is able to maintain the throughput of a baseline monolithic Linux system.

Future work will extend \sys{} to enforce protection against
return- and jump-oriented attacks using Control-flow Enforcement
Technology (CET)~\cite{intel_sdm}. Other research directions include more fine-grained
isolation of dynamically allocated objects, boot-time compartment initialization, and dynamic compartment creation.


\bibliographystyle{ACM-Reference-Format}
\bibliography{references}

\end{document}